\documentclass[10pt,conference,a4]{IEEEtran}

\usepackage{amsmath,dsfont}
\usepackage{amsthm}
\usepackage{amsfonts}
\usepackage{amssymb}
\usepackage{graphicx}
\usepackage{cite}
\usepackage{subfigure}
\usepackage{balance}
\usepackage{url}
\usepackage{hyperref}
\usepackage{multirow} 
\usepackage{algorithm}
\usepackage{algorithmic}
\usepackage[short]{optidef}
\usepackage[symbol]{footmisc}

\usepackage{blindtext}

\ifodd 1
\usepackage{soul}
\usepackage{color}
\setstcolor{red}

\definecolor{purple}{rgb}{0.0, 0.6, 0.9}
\definecolor{pink}{rgb}{0.96, 0.76, 0.76}

\newcommand{\del}[1]{\st{#1}} 

\newcommand{\com}[1]{\textbf{\color{red} (Mian: #1)}} 
\newcommand{\response}[1]{{\color{magenta} (Response: #1)}} 

\else

\newcommand{\del}[1]{}

\newcommand{\com}[1]{}
\newcommand{\comg}[1]{}
\newcommand{\response}[1]{}
\fi

\definecolor{morange}{rgb}{0.8,0.2,0}
\definecolor{mblue}{rgb}{0,0.1,0.8}
\definecolor{mgreen}{rgb}{0,0.8,0.1}
\definecolor{mred}{rgb}{1,0,0}

\definecolor{myhlcolor}{rgb}{0.91,0.94,0.62}
\sethlcolor{myhlcolor}

\usepackage{array}
\newcolumntype{L}[1]{>{\raggedright\let\newline\\\arraybackslash\hspace{0pt}}m{#1}}
\newcolumntype{C}[1]{>{\centering\let\newline\\\arraybackslash\hspace{0pt}}m{#1}}
\newcolumntype{R}[1]{>{\raggedleft\let\newline\\\arraybackslash\hspace{0pt}}m{#1}}

\makeatletter
\renewcommand\st[1]{\@bsphack\@esphack}
\makeatother

\begin{document}
\title{Collaborative Edge Learning in MIMO-NOMA Uplink Transmission Environment}

\author{\IEEEauthorblockN{Mian~Guo\IEEEauthorrefmark{1},~Chun~Shan\IEEEauthorrefmark{1}, ~Mithun~Mukherjee\IEEEauthorrefmark{2},~Jaime~Lloret\IEEEauthorrefmark{3}, Quansheng~Guan\IEEEauthorrefmark{4},
} \\

\IEEEauthorblockA{\IEEEauthorrefmark{1}School of Electronics and Information, Guangdong Polytechnic Normal University, P.R. China}
\IEEEauthorblockA{\IEEEauthorrefmark{2}School of Artificial Intelligence, Nanjing University of Information Science and Technology, P.R. China}
\IEEEauthorblockA{\IEEEauthorrefmark{3}Instituto de Investigación para la Gestión Integrada de Zonas Costeras (IGIC), Universitat Politecnica de Valencia, Spain}
\IEEEauthorblockA{\IEEEauthorrefmark{4}School of Electronic and Information Engineering, South China University of Technology, P.R. China}}

\maketitle

\begin{abstract}
Multiple-input multiple-output non-orthogonal multiple access (MIMO-NOMA)  cellular network is promising for supporting massive connectivity. This paper exploits low-latency machine learning in the MIMO-NOMA uplink transmission environment, where a substantial amount of data must be uploaded from multiple data sources to a  one-hop away edge server for machine learning. A delay-aware edge learning framework with the collaboration of data sources, the edge server, and the base station, referred to as \texttt{DACEL}, is proposed. Based on the delay analysis of \texttt{DACEL}, a  NOMA channel allocation algorithm is further designed to minimize the learning delay. The simulation results show that the proposed algorithm outperforms the baseline schemes in terms of learning delay reduction. 
\end{abstract}

\section{Introduction}\label{sect: intr}

\IEEEPARstart{W}{ith}  the development of the fifth generation (5G)  \st{and beyond} cellular network,  there will be new  opportunities for various advanced applications in  Internet of Things (IoT) and vehicle-to-everything (V2X) areas \cite{Boccuzzi2019}. Generally, IoT and V2X applications are data-driven and delay sensitive, such as autonomous driving and real-time manufacturing\st{ \cite{Porambage_2018}}, which desire efficient and effective machine learning for data analytics with low latency.

Mobile edge computing (MEC) is a promising paradigm for delay-sensitive machine learning \cite{Shi_2016,GuoM2020}.  By deploying machine learning servers at the network edge closer to data sources (e.g., IoT devices), the network transmission delay for uploading data from data sources to a remote server could be explicitly reduced.  However, due to limited wireless resource, the data collection delay, which is defined as the duration from the time that  the first data source starts to upload the first data  to the time that the edge server receives all data from all data sources, would be non-negligible. For example, under  orthogonal multiple access (OMA) techniques, \st{such as time division multiple access (TDMA), frequency division multiple access (FDMA), and code division multiple access (CDMA),}  due to the limited number of orthogonal uplink resources,  all IoT devices cannot simultaneously upload their data to the edge server. As a result, the  waiting time for the data collection would dramatically degrade the learning efficiency.

Recently, non-orthogonal multiple access (NOMA)\st{, which is regarded as a promising multiple-access technique for 5G and beyond cellular systems,}  has gained ever-growing attention for supporting massive connectivity and the increasing demands of advanced IoT applications~\cite{islam2019nonorthogonal, DingZ_2015, LiC2021}.  In NOMA, multiple users are allowed to use the same resource block. Thus, a massive number of IoT devices are possible to  simultaneously upload their data to an edge server, which may reduce the data collection delay.  However, in contrast to OMA, users under NOMA would experience inter-user interference\st{ because of the non-orthogonal resource allocation}. Besides, the more users use the same NOMA channel, the more complex for the receiver to separate different users' signals. Accordingly,  multiple-input multiple-output (MIMO)-NOMA uplink transmission  \cite{islam2019nonorthogonal} is promising for edge  learning because of  multiple NOMA channels. IoT devices can be distributed to different NOMA channels to reduce the intra-cell interference and lower the  receiver's algorithm complexity. \emph{However, the heterogeneity of dataset sizes from various IoT devices and the nature that user channels are in the form of matrices are the major challenges for  edge learning in a MIMO-NOMA uplink environment}. 

In this paper, our objective is to design a low-latency edge learning framework, referred to  as \texttt{DACEL}, in  a MIMO-NOMA uplink transmission environment. To support simultaneous data upload, we use base station with the collaboration of edge servers and IoT devices to allocate NOMA uplink resources to IoT devices (data sources). We also set a delay deadline for the data collection step of a learning process, aiming to avoid long waiting time for the data collection from worse sources/uplinks.  By analyzing the delay of \texttt{DACEL},  a NOMA channel allocation algorithm is further proposed to achieve the goal of minimizing the edge learning delay.

\section{ Delay-Aware Collaborative Edge Learning }\label{sect:mod_desc}

\subsection{System Model}\label{subsect:sys_mod}
\begin{figure}[t]
\centering{}
\includegraphics[width=0.45\textwidth]{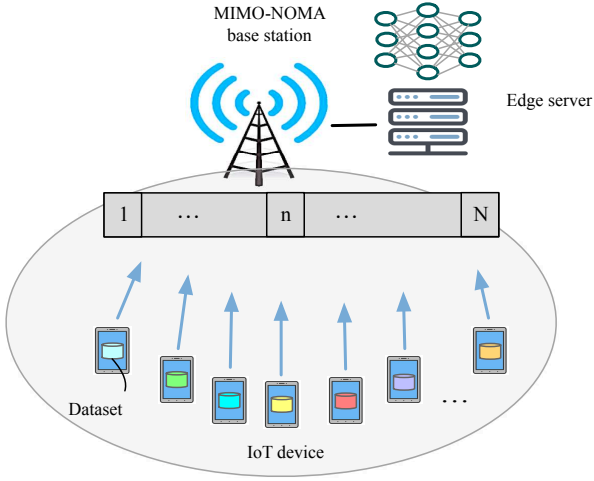}
\caption{An edge learning model with MIMO-NOMA.}\label{fig:sys_mod}
\vspace*{-6pt}
\end{figure}

As illustrated in Fig. \ref{fig:sys_mod}, we consider a MEC system with multiple IoT devices and an edge server. The IoT devices are randomly distributed in a MIMO-NOMA wireless network, while the edge server is located at the center of the network (e.g., integrated in the base station). The base station manages the communications between the IoT devices and the edge server. The machine learning algorithm is assumed to be deployed at the edge server, while the datasets are from  IoT devices. In special, we consider the machine learning for  delay-sensitive and computation intensive applications that need heterogeneous datasets from multiple IoT devices, such as  autonomous driving and real-time manufacturing. We consider a synchronous learning model, that is, the training will not start until the edge server has aggregated the datasets. 

\st{Specifically, we assume that at a learning process, firstly,  the IoT devices  send their up-to-date datasets to the edge server via the wireless network, where a dataset comprises  a number of up-to-date data samples from the  sensing environments. Then,  the edge server aggregates the datasets and  performs  training, aiming at supporting IoT applications timely and efficiently. Take real-time manufacturing as an example, cameras from different locations and different angles of a workspace record the environments they see and will send the photo/video samples to the edge server. The edge server learns from this  data via machine learning and then sends  commands to control the actions of industrial robots on-demand. }

We denote $\mathcal{M}=\{1, 2, \ldots, M\}$ as the set of IoT device in MEC system. We further denote $L=(L_1, L_2, \ldots, L_m, \ldots, L_M)$ as the vector of dataset size (in bits), where $L_m$ is the dataset size of IoT device, $m, \forall m \in \mathcal{M}$. Moreover, we denote $W=(W_1, W_2, \ldots, W_m, \ldots, W_M)$ as the corresponding computing requirement vector, where $W_m$ indicates that the dataset from the IoT device, $m$, has $W_m$ number of samples.  

\subsection{MIMO-NOMA Uplink}\label{subsect:link_mod}
As illustrated in Fig. \ref{fig:sys_mod}, we consider a MIMO-NOMA uplink transmission environment\cite{ZengM_2019}, where the base station communicates with IoT devices (assume that the communication between the base station and the edge server can be ignored due to their closer locations). The base station is equipped with $N$ antennas, while all IoT devices are equipped with single antennas. Thus, there are $N$ number of channels available to the IoT devices, while an IoT device can only use one of these channels to upload its dataset to the edge server.  Besides, IoT devices use the same channel will be supported by the same beamforming vector. We assume that the full channel state information (CSI) is available at the base station.

 Let $\mathcal{N}=\{1, 2, \ldots, N\}$ be the available  channel set.   Let $\mathcal{I}=\left(\mathcal{I}_1,\mathcal{I}_2, \ldots, \mathcal{I}_m, \ldots, \mathcal{I}_M\right)$ be the channel decision vector, where $\mathcal{I}_m=\left(I_{m, 1}, I_{m, 2} \ldots, I_{m, n}, \ldots, I_{m, N} \right)$ is the channel decision vector of IoT device $m$, in which $I_{m, n}=\{0, 1\}$  for $m\in\mathcal{M}$ and $n\in\mathcal{N}$ is a binary channel allocation variable for IoT device $m$ in channel $n$. That is, if  channel $n$ is allocated to IoT device $m$, then, $I_{m, n}=1$, otherwise, $I_{m, n}=0$.

\subsection{\texttt{DACEL} Framework}\label{subsect:dacel}
We propose \emph{delay aware collaborative edeg learning}, namely \texttt{DACEL}, in \textit{Framework} \ref{alg: noma_learning} to perform machine learning in MIMO-NOMA uplink base MEC  with the collaboration of IoT devices, the  edge server and the base station. As shown in  \textit{Framework} \ref{alg: noma_learning}, under \texttt{DACEL},  a machine learning process mainly experiences the following steps. At the beginning, the edge server sends an information of  \emph{data collection request} to IoT devices, which requires the IoT devices to upload their up-to-date datasets to the edge server for model training.  Due to dynamic wireless environments, some feedbacks may be particularly long. Therefore, in order to avoid extremely long waiting time, a \emph{timer} is used to count the waiting time.

\floatname{algorithm}{Framework}
\begin{algorithm}[t]
\caption{\small Delay-aware collaborative edge learning}
\label{alg: noma_learning}
\begin{enumerate}
\item[1: ] \texttt{Initialization:} The edge server sends \emph{ data collection request} to the IoT device set $\mathcal{M}$, and initiates a \emph{timer} to count the waiting time.

\item[2: ] \texttt{Wirelss resource request:} Each IoT device $m\in\mathcal{M}$ sends  \emph{channel request} to the base station, along with its dataset size $L_m$.

\item[3: ] \texttt{NOMA channel allocation:} The base station initiates \emph{channel allocation} algorithm to obtain $\mathcal{I}^{*}=\left(\mathcal{I}_1,\mathcal{I}_2, \ldots, \mathcal{I}_m, \ldots, \mathcal{I}_M\right)$, and then feeds back to IoT devices. 
 
\item[4: ] \texttt{Dataset uploading:} Each  IoT device $m\in\mathcal{M}$ sends its dataset to the edge server via the allocated channel (e.g., via the channel $n^{'}$ where $I_{m, n^{'}}=1$). 

\item[5: ] \texttt{Dataset aggregation:} When the edge server receives all the datasets, or, the \emph{timer} reaches its delay deadline, it aggregates the received datasets.

\item[6: ] \texttt{Training:} The edge server trains a learning model based on the aggregated datasets until the model achieves at a desired performance or interrupted. 

\item[7: ] \texttt{Result feedback:} The learning result  feeds back to the IoT application. 

\end{enumerate}
\end{algorithm}

When IoT devices receive the request, they will send \emph{channel request} to the base station along with its dataset size. Then, the base station will initiate a channel allocation algorithm to allocate channels to IoT devices considering the CSI and the dataset sizes, aiming at reducing the data collection delay. After that, the IoT device will upload the dataset to the edge server via the   allocated channel. After received all the datasets, or the \emph{timer} reaches its delay deadline, the edge server aggregates the received datasets. Note that, we assume that the server will drop the dataset arrived after the delay deadline of the \emph{timer}. Then, the edge server starts to train a learning model based on the aggregated datasets. The training will be stopped when the model achieves at a desired performance or the number of rounds  reaches the pre-setting.  Finally, the learning result will be feedback to the IoT application, which ends the learning. 

\subsection{Delay Analysis}\label{subsect:delay_mod}
As illustrated in  \textit{Framework} \ref{alg: noma_learning}, the delay of a machine learning process can be defined as a duration of  the time from the edge server sends the \emph{data collection request} to the time that the learning result feeds back to the application.  Comparing to the time duration of other steps in \texttt{DACEL}, the  time durations of both \texttt{dataset uploading} and \texttt{training} dominate the delay of a learning process. Accordingly,  we mainly consider the  delay in the \texttt{dataset uploading} and \texttt{training} steps. 

\subsubsection{Uploading Delay}
Assume that the bandwidth of channels are identical.  Let $B$ be the bandwidth of a channel. Let $h_{m, n}$ be the channel power gain of IoT device $m$ at channel $n$. Let $S_{n}$ be the set of IoT devices sharing the $n$th channel, where $n\in\mathcal{N}$. Then, the uplink transmission rate from IoT device $m$ to the base station via channel $n$ is derived by
\begin{equation}\label{eq:rate_def}
R_{m, n}=B log_2\left( 1+\frac{I_{m, n} P_{m, n} h_{m, n}}{\mathcal{G}_{m, n^{-}}+\gamma_{m, n} + N_0}        \right),
\end{equation}
where $P_{m, n}$ is the transmission power; $\mathcal{G}_{m, n^{-}}$ is the inter-cell interference from the users with lower channel power gain  in all other channels \cite{ZengM_2019}; $\gamma_{m, n}$ is the intra-cell interference; $N_{0}$ is the noise power.   

The inter-cell interference $\mathcal{G}_{m, n^{-}}$ is derived by
\begin{equation}
\mathcal{G}_{m, n^{-}}\!\!=\!\! \sum_{k\in\mathcal{M}\setminus\{m\}} \sum_{n^{'}\in \mathcal{N}\setminus\{n\}} \!\! I_{k, n^{'}} \mathds{1} \left( h_{k, n^{'}} < h_{m, n} \right) P_{k, n^{'}} h_{k, n^{'}},
\end{equation}
where $\mathds{1} (K)=\{0, 1\}$, if $K$ is true then, $\mathds{1}(K)=1$ and $\mathds{1}(K)=0$, otherwise.

Under NOMA, a user will also experience intra-cell interference from all the users with lower channel power gain \cite{Tabassum_2017}. Thus, the intra-cell interference is derived by 
\begin{equation}\label{eq:interference_def}
\gamma_{m, n}=\sum_{k\in S_n} I_{k, n} \mathds{1} \left( h_{k, n} < h_{m, n} \right) P_{k, n} h_{k, n}.
\end{equation}
Accordingly, the uplink transmission rate of IoT device $m$ can be expressed by $R_{m}=\sum_{n\in\mathcal{N}} R_{m, n}$.
Therefore, the dataset uploading delay is derived by
\begin{equation}\label{def:upload_delay}
D_{m}=\frac{L_m}{R_{m}}=\frac{L_m}{\sum_{n\in\mathcal{N}} R_{m, n}} \:.
\end{equation}
We assume that the IoT devices send their datasets almost at the same time. As shown in \textit{Framework} \ref{alg: noma_learning}, the \texttt{dataset uploading} step finishes when the edge server receives all the datasets, or the \emph{timer} reaches its delay deadline. Therefore, the dataset uploading delay of the system can be expressed by
\begin{equation}\label{eq:max_upload_delay}
D_{\text{Tx}}=\min\left[\max_{m \in \mathcal{M}}  D_{m}, D_{\text{Tx}}^{\text{Max}}\right],
\end{equation}
where $D_{\text{Tx}}^{\text{Max}}$ is the maximum delay that the edge server can tolerate for data collection. 

\subsubsection{Training Delay}\label{subsubsect:training_delay}
Denote $S_{\text{B}}$ and $F_{\text{CPU}}$ as the batch size  of the machine learning algorithm and the computing rate, respectively. Then, the delay of the training with a batch size can be derived by $D_{\text{CPU, B}} = S_{\text{B}}/F_{\text{CPU}}$.
Thus, the delay of a round of training can be derived by
\begin{equation}\label{eq:D_cpuR}
D_{\text{CPU, R}}=D_{\text{CPU, B}} \: \frac{\sum_{m\in\mathcal{M}}  \mathds{1}(D_m \leq D_{\text{Tx}}^{\text{Max}}) W_{m}}{S_{\text{B}}},
\end{equation}
where $\sum_{m\in\mathcal{M}}  \mathds{1}(D_m \leq D_{\text{Tx}}^{\text{Max}}) W_{m}$ is the sum of data received at the  edge server. 

We assume that, after $K_{\text{R}}$ number of rounds, the training is converged to the performance objective, or, we will interrupt the training after $K_{\text{R}}$ number of rounds. Then, the total delay of a training process is calculated by
\begin{equation}\label{eq:D_cpu}
D_{\text{CPU}} = K_{\text{R}} D_{\text{CPU, R}} \:.
\end{equation}
According to \cite{McMahanMRA_16, YangQ_2019}, the number of rounds requiring to run for model accuracy can be expressed by 
$K_{\text{R}} = \alpha/\left(\sum_{m\in\mathcal{M}}  \mathds{1}(D_m \leq D_{\text{Tx}}^{\text{Max}})  W_m\right)^{\sigma} $, where $\alpha > 0$ and $\sigma > 1$ are factor parameters.

Therefore, the total learning delay is expressed as 
\begin{equation}\label{eq:D}
D=D_{\text{Tx}} + D_{\text{CPU}}.
\end{equation}

\section{Algorithm for NOMA Channel Allocation }\label{sect:dd_maxH}
\subsection{Problem Formulation}\label{subsect:pro}
The delay analysis in Section \ref{subsect:delay_mod} indicates that, the channel allocation decision $\mathcal{I}$ affects the dataset uploading  and training delays, as illustrated in \eqref{eq:rate_def} and \eqref{eq:D_cpuR}, respectively, thus affects the learning delay. Accordingly,  in the \texttt{NOMA channel allocation} step, our  objective is to find optimal channel allocation policy to minimize the learning delay. We formulate the above problem as the NOMA channel allocation (NOMA-CA) for delay-minimization  problem and write as
\begin{mini!}[2]
{\mathcal{I}^{*}}{D }{}{}{\label{eq:optimization0}}
\addConstraint{\sum_{n\in\mathcal{N}} I_{m, n}=1, \forall m\in\mathcal{M}    }{}{}{\label{eq:optimization3}}
\addConstraint{ \eqref{eq:rate_def}-\eqref{eq:D}.}{}{}{\label{eq:optimization5}}
\end{mini!}
Note that, for simplification, this paper considers that one IoT device will use at most one channel (e.g., equipped with a single antenna) to transmit its dataset to the edge server. Thus,~\eqref{eq:optimization3} holds for all $m\in\mathcal{M}$. 

\subsection{DD-maxH }\label{subsect:dd_maxH}
As described in \eqref{eq:max_upload_delay} and \eqref{eq:D_cpu}, the more number of datasets are successfully uploaded  within the delay deadline $D_{\text{Tx}}^{\text{Max}}$, the smaller number of rounds are required to train. In addition,  the smaller of $\max_{m\in\mathcal{M}} D_m$, the lower delay of a learning. Therefore, we design a NOMA channel allocation algorithm, namely, \emph{delay descent max-H} (\emph{DD-maxH}), to reduce the maximum dataset uploading delay of all IoT devices, aiming at achieving the goal of minimizing the learning delay. 

As illustrated in \textit{Algorithm} \ref{alg: maxmin}, the algorithm mainly consists of two steps, including \emph{max-H decision} and \emph{delay descent decision update}. In the \emph{max-H decision} step, each IoT device is  allocated a channel that has the best channel quality. Since different IoT devices may experience different inter-cell and intra-cell interferences according to the channel decisions, some IoT device may experience large dataset uploading delay, while others may experience smaller delays.  \st{The learning delay as well as the learning accuracy is significantly affected by the maximum uploading delay as shown in \eqref{eq:max_upload_delay} and \eqref{eq:D_cpu}.} Thus, in  the  \emph{delay descent decision update} step, we will reduce the maximum dataset uploading delay  by adjusting the channel allocation decision of the IoT device that will experience the maximum dataset uploading delay.

\floatname{algorithm}{Algorithm}
\begin{algorithm}
\caption{\small Delay descent max-H (DD-maxH) algorithm }
\label{alg: maxmin}
\textbf{Input:} $\mathcal{M}$, $\mathcal{N}$, $L_m$, $P_m$ for $m\in\mathcal{M}$, $B$, $h_{m, n}$ for  $m\in\mathcal{M}$ and $n\in\mathcal{N}$,   $N_0$.

\textbf{Output:} $I$.

\vspace*{0.5em}

 \textbf{Initialization: }  $I=\{\}$.

  \textbf{Max-H decision: }

 \hspace*{0.5em} \textbf{For} $m \in \mathcal{M}$, \textbf{do}
\begin{enumerate}
\item Find a subchannel $n^{'}$ that satisfies:  

$n^{'}=\text{argmax}_{n\in\mathcal{N}} h_{m, n}$.

\item Set $I_{m, n^{'}}=1$.

\end{enumerate}

 \textbf{Delay descent decision update:} 

\begin{enumerate}
\item[a)] Calculate $D_{m}$ for all $m\in\mathcal{M}$  with \eqref{def:upload_delay}.

\item[b)] Find $m^{'}$ that satisfies: $m^{'}=\text{argmax}_{m\in\mathcal{M}} D_{m}$.

\item[c)] Find $D_{m^{'}, n^{''}}$ that satisfies: 
\begin{equation}
D_{m^{'}, n^{''}}=\text{min}_{n\in\mathcal{N}\setminus \{n^{'}\}} D_{m^{'}, n}, 
\end{equation}
where  $n^{'}$ is the channel that presently satisfies $I_{m^{'}, n^{'}}=1$. $D_{m^{'}, n}=L_{m^{'}} / R_{m^{'}, n}$ is the dataset uploading delay if channel $n$ is selected (equivalently, $I_{m^{'}, n^{'}}=0$ and $I_{m^{'}, n}=1$.), where  $R_{m^{'}, n}$ is derived by \eqref{eq:rate_def}.

\item[d)] \textbf{If} $D_{m^{'}, n^{''}} < D_{m^{'}, n^{'}}$, \textbf{do}

\begin{enumerate}
\item[(1)] Set $I_{m^{'}, n^{'}}=0$ and $I_{m^{'}, n^{''}}=1$.
\item[(2)] Go to step a).
\end{enumerate}

\end{enumerate}

\end{algorithm}

Specifically, as shown in  \textit{Algorithm} \ref{alg: maxmin}, in the \emph{delay descent decision update} step, we first calculate the dataset uploading delay of every IoT device based on the present channel allocation decision. Then, we find the IoT device with the maximum uploading delay. We estimate whether there exist other channel allocation decisions that can reduce the maximum uploading delay. If there exists, then we switch the decision to the new channel that will reduce the maximum uploading delay. The \emph{delay descent decision update} step repeats until no further channel switching can reduce the maximum uploading delay.  Then, the final channel allocation decision is the best decision that achieves the goal of minimizing the learning delay.

\section{Performance Evaluation}\label{sect:sim}

This section investigates the  delay  performance of the proposed \emph{DD-maxH} NOMA channel allocation algorithm under the  \texttt{DACEL} framework. In each of the simulation run, the  MEC system has a  computation task of object classification, which will be run at the edge server with   independent image datasets from distributed IoT devices. Each dataset is a subset of MNIST\cite{mnist}, which comprises 60,000 training images and 10,000 testing images with 10 object classes (i.e., 10 digits). An image is equivalent to a sample.  
The basic parameter settings  are listed in Table \ref{Tb:parameter}. 

\begin{table}[t]
\caption{The basic parameter settings}\label{Tb:parameter}
\vspace*{-\baselineskip}
\centering\begin{tabular}{|l|c||c|}
\hline
\multicolumn{2}{|c||}{Parameters}&Value \\
\hline
IoT & Number of IoT devices, $M$ & 6\\
\cline{2-3}
devices &Maximum power, $P^{\text{Max}}$ (W)&0.1\\
\hline
Dataset&\!\!Mean dataset size, $L_m$ (MB/dataset)\!\!& 1.8\\
\cline{2-3}
\!\!and training\!\!& Mean, $W_m$ (Samples/dataset)  & $8.0\times10^{3}$\\
\cline{2-3}
 model&Batch size (Samples/batch) &100\\
\hline
\!\!Edge server\!\!&Computing capability (Samples/s) & $1.0\times10^{3}$\\
\hline
Uplink &Number of channels, $N$  & 3\\
\cline{2-3}
&Channel bandwith, $B$ (MHz) & 5.0\\
\cline{2-3}
 info&Channel power gain, $h$ &[$4\times10^{-10}$, 0.035]\\
\cline{2-3}
&Noise, $N_0$ & $10^{-10}$\\
\hline 
\end{tabular}
\end{table}

We compare the proposed algorithm, e.g., \emph{DD-maxH}, with three \st{benchmarked} baseline algorithms, e.g., \emph{max-H-greedy},  \emph{min-H-greedy}, and \emph{strictly-priority-delay-minimization} (namely, \emph{SPDM}). In \emph{max-H-greedy}, for every IoT device, the channel that has the best channel quality\st{ (e.g., the channel power gain of that IoT device in that channel is the highest among all channels that the IoT device can access)}, that is, channel $n$ that satisfies $n=\text{argmax}_{k\in\mathcal{N}} h_{m, k}$, is allocated to IoT device $m$. By contrast, under \emph{min-H-greedy}, the channel $n$ that satisfies $n=\text{argmin}_{k\in\mathcal{N}} h_{m, k}$ is allocated to IoT device $m$ for $m\in\mathcal{M}$. With \emph{min-H-greedy}, IoT devices may experience less interference under the MIMO-NOMA uplink environments. 
In \emph{SPDM}, the IoT devices are allocated channel resource one-by-one in the descent order of their dataset sizes. For example, when an IoT device $m\in\mathcal{M}$ is the present  un-allocated device that has the largest dataset size,  the channel $n$ that will provide the minimum dataset uploading delay $D_{m, n}=min_{n^{'}\in\mathcal{N}} D_{m, n^{'}}$ would be allocated to this device. 
 
 We evaluate the performance of the above algorithms by setting the \emph{timer}, equivalently, the dataset uploading delay deadline of all the investigated algorithms as $D_{\text{Tx}}^{\text{Max}}=\text{max}_{m\in\mathcal{M}} D_m^{\text{max-H-greedy}}$, where  $D_m^{\text{max-H-greedy}}$ is the uploading delay of the dataset from IoT device $m$ for $m\in\mathcal{M}$ under the \emph{max-H-greedy} algorithm.

\subsection{Adaptive to IoT Devices}\label{subsect:varying_client}
\begin{figure}
\centering
\includegraphics[width=0.43\textwidth]{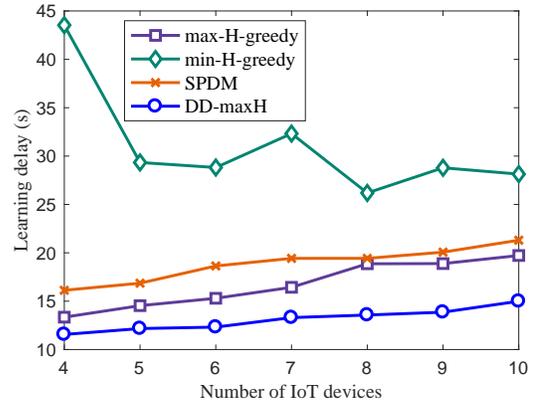}
\centering
\caption{The impact of   IoT devices on the learning delay.}
\label{fig:delay_varyingM_Dth_maxH}
\vspace*{-6pt}
\end{figure}

This scenario observes the performance by varying  the number of IoT devices, equivalently, varying the sum of dataset sizes.  We  observe the  delay performance by varying $M$ from 4 to 10. As illustrated in Fig. \ref{fig:delay_varyingM_Dth_maxH}, under all the investigated algorithms, the learning delay increases with the increasing number of IoT devices. This is because, the aggregated dataset size increases with the increasing number of IoT devices. Thus, the training delay increases. 

Our proposed \emph{DD-maxH} outperforms the others by providing the lowest learning delay under various number of IoT devices, as shown in Fig. \ref{fig:delay_varyingM_Dth_maxH}. This is because,  by iteratively reducing the maximum uploading delay via adjusting the channel decision, the \emph{DD-maxH} algorithm provides the lowest maximum uploading delay while successfully uploading the most amounts of samples to the edge server. Thus, both uploading delay and training delay given by \emph{DD-maxH}  are the lowest. \st{, as shown in Figs. \ref{fig:upload_delay_varyingM_Dth_maxH}-\ref{fig:train_delay_varyingM_Dth_maxH}.}

\st{Note that, under both \emph{min-H-greedy} and \emph{SPDM}, since the maximum dataset uploading delay would exceed the uploading delay deadline $D_{\text{Tx}}^{\text{Max}}$, some datasets cannot be successfully uploaded to the edge server, thus the edge server is required to train with more number of rounds for model accuracy. Therefore, it is unsurprising that the training delay given by both  \emph{min-H-greedy} and \emph{SPDM} are explicitly higher than those given by \emph{max-H-greedy} and \emph{DD-maxH}, as shown in Fig. \ref{fig:train_delay_varyingM_Dth_maxH}.}

\subsection{Adaptive to Channels}
\begin{figure}[t]
\centering{}
\includegraphics[width=0.43\textwidth]{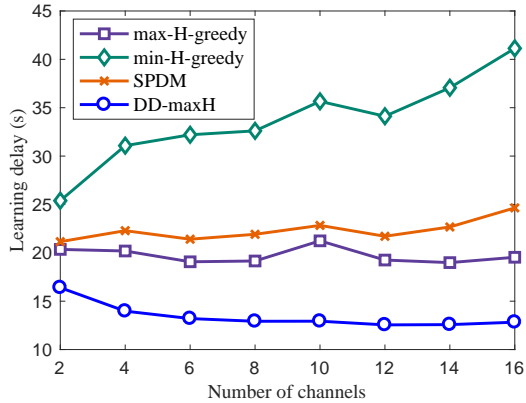}
\caption{The effect of available channels on the learning delay.}\label{fig:delay_varyingN_Dth_maxH}
\vspace*{-6pt}
\end{figure}

In this scenario, we observe the performance by varying the number of available NOMA channels from 2 to 16. We set the number of IoT devices to $M=10$. Other parameters are listed in Table \ref{Tb:parameter}. As illustrated in Fig. \ref{fig:delay_varyingN_Dth_maxH}, the learning delays given by both \emph{max-H-greedy} and \emph{SPDM} vary less under various  number of channels. This is because, under \emph{max-H-greedy}, the channel that has the highest channel power gain is allocated to an IoT device.  Similarly, under \emph{SPDM}, the channel that has the highest channel power gain  will be allocated to an IoT device with high probability for providing  a lower dataset uploading delay to that IoT device. Although the value of the highest channel power gain may increase with the increasing number of channels, the inter-cell and intra-cell interferences would also increase with the increasing number of channels. Thus, the uplink transmission rate given by both  \emph{max-H-greedy} and \emph{SPDM} may not increase. Therefore, the number of channels has little effect on the channel allocation decision under  both \emph{max-H-greedy} and \emph{SPDM}.

\begin{table}[t]
	\caption{Model accuracy (\%)}\label{Tb:mod_accuracy}
	\vspace*{-\baselineskip}
	\centering\begin{tabular}{|l|c|c|c|c|}
		\hline
		M&\emph{max-H-greedy}&\emph{min-H-greedy}&\emph{SPDM} & \emph{DD-maxH}\\
		\hline
		4&91.14&90.45 & 90.91&91.14\\
		\hline
		5&91.50&90.61&91.18&91.51\\
		\hline
		6&91.76&90.95&91.52&91.78\\
		\hline
		7&91.84&91.25&91.66&91.90\\
		\hline
		8&91.85&91.40&91.90&91.95\\
		\hline
		9&91.92&91.53&92.01&91.98\\
		\hline
		10&91.68&91.73&92.02&91.80\\ 
		\hline 
	\end{tabular}
\vspace{-8pt}
\end{table}

On contrary, the learning delay given by \emph{min-H-greedy} increases with the increasing number of channels, as illustrated in Fig. \ref{fig:delay_varyingN_Dth_maxH}. Under \emph{min-H-greedy}, the channel that has the lowest channel power gain is allocated to an IoT device. Although this behavior can reduce the intra-cell interference,  the uplink transmission rate would decrease with the increasing number of channels. This is because  the value of the  lowest channel power gain reduces with the increasing number of channels.

The efficiency of the proposed \emph{DD-maxH} algorithm is again illustrated by providing the lowest learning delay under various number of channels, as shown in Fig. \ref{fig:delay_varyingN_Dth_maxH}. Different from the other investigated algorithms, under \emph{DD-maxH}, the learning delay decreases with the increasing number of IoT devices. This is because, with more number of channels available,  the maximum dataset uploading delay could be reduced more  by selecting optimal channels from more wide range of channels.

\subsection{Model Accuracy}
Finally, we list the model accuracy of the investigated algorithms under a varying number of IoT devices. As shown in Table \ref{Tb:mod_accuracy}, the investigated algorithms provide similar model accuracy with a different number of IoT devices, which demonstrates the accuracy of training model described in~\eqref{eq:D_cpu}.

\section{Conclusions}\label{sect:con}
In this paper, we have studied a collaborative edge learning problem in a  MIMO-NOMA uplink transmission environment. A \texttt{DACEL} framework has been designed to support machine learning at the network edge with the collaboration of edge servers, IoT devices, and base station. We have further proposed a MIMO-NOMA channel allocation algorithm to reduce the learning delay of \texttt{DACEL}.  Finally, the simulation results have illustrated the efficiency of the proposed method for learning delay reduction while satisfying the model accuracy requirement.  

\section*{Acknowledgment}
This work was supported in part by the Key-Area Research and Development Program of Guangdong Province under Grant 2020B0101130023, National Natural Science Foundation of China under Grant 61901128, Startup Foundation for Introducing Talent of NUIST (1521632101005), National Key R\&D Program of China--Industrial Internet Application Demonstration-Sub-topic Intelligent Network Operation and Security Protection (2018YFB1802402), Project supported by Innovation Group Project of Southern Marine Science and Engineering Guangdong Laboratory Zhuhai  (311021011). The corresponding author is Mithun Mukherjee.


\end{document}